\documentstyle[amssymb,a4,12pt]{article}

\input tcilatex
\begin{document}

\title{SOME COMMENTS ON THE RHS FORMULATION OF RESONANCE SCATTERING.}
\author{M. Gadella$^1$ and A. R. Ord\'{o}\~{n}ez$^2$ \and \bigskip \\
1 Departamento de F\'{\i}sica Te\'{o}rica. Facultad de Ciencias,\\
c. Real de Burgos s.n., 47011 Valladolid, Spain.\\
2 Facultad de Ciencias Exactas, Ingenier\'{\i}a y \\
Agrimensura. Universidad Nacional de Rosario, \\
Av. Pellegrini 250. (2000) Rosario, Argentina.}
\maketitle

\begin{abstract}
We discuss the validity of a formula concerning a relation between
functionals in quantum resonance scattering, which is often used in the
current literature.
\end{abstract}

\section{Introduction.}

This paper is a contribution to the theory of resonance scattering in which
we discuss the validity of some formulas and concepts that appear in the
current literature. This kind of formulas are usually derived formally and
used directly. Thus, an interpreation of them from the point of view of
mathematical rigor is usually necessary to know their conditions of validity.

By resonance scattering we mean a scattering process that produces
resonances \cite{1}, \cite{2}, \cite{3}. We assume that this scattering
process satisfies properties of regularity such as: existence of the M\o
ller opeartors, asymptotic completeness, absence of singular continuous
spectrum, etc. Resonances are characterized by pair of poles on the analytic
continuation of the $S$-matrix beyond the cut on the positive semiaxis in
the energy representation \cite{1}, \cite{3}, \cite{4}. These pairs of poles
have a real part which coincides with the resonance energy and an imaginary
part which is the inverse of the mean life. They are complex conjugate of
each other. Some additional conditions on the behaviour of the $S$-matrix on
the second sheet are also imposed \cite{5}. Under this conditions, we can
separate the exponentially decaying part of a resonance from the background 
\cite{2}, \cite{5}. This exponentially decaying part, also called Gamov
vector, is not a regular state in a Hilbert space and can be defined as a
functional belonging to an extension of Hilbert space. This extension is
given by a rigged Hilbert space (RHS).

Gamov vectors can be defined even if regularity conditions for the $S$%
-matrix are not present, but then, the difficulty arises in the separation
of the exponentially decaying part from the background. If the $S$-matrix
poles are not simple, in addition to the exponentially decaying Gamov
vectors, it must exist other functionals, also called Gamov vectors, for
which the decay is a product of an exponential contribution times a
polynomial on the time. These objects are the so called multiple pole Gamov
vectors, which are introduced elesewhere \cite{6}, \cite{7}. Their energy
representation has a Breit-Wigner part (as the exponentially decaying state
vectors) plus other contributions \cite{1}.

The RHS formalism for resonances and Gamov vectors has been introduced many
times \cite{1}, \cite{2}, \cite{5}, \cite{6}, \cite{7}, \cite{8}, $\cite{9}$%
. Therefore, we refer the interested reader to the mentioned literature and
the references quoted therein for this formalism. Here, we are using the
notation of reference \cite{7}. See also \cite{10}. Nevertheless, some
remainder is necessary in order to make this paper self contained.

We start with the space of Hardy functions on the upper, ${\cal H}_{+}^2$,
an lower, ${\cal H}_{-}^2$, half planes \cite{11} and we intersect them with
the Schwartz space $S$. The restrictions of the functions on these spaces to
the positive semiaxis give two locally convex nuclear spaces $\cite{5}$,
which are denoted as $\left. S\cap {\cal H}_{\pm }^2\right| _{{\Bbb R}^{+}}$
and their respective duals by $^{\times }\left( \left. S\cap {\cal H}_{\pm
}^2\right| _{{\Bbb R}^{+}}\right) $. The triplets given by

\begin{equation}
\left. S\cap {\cal H}_{\pm }^2\right| _{{\Bbb R}^{+}}\,\subset \,L^2({\Bbb R}%
^{+})\,\subset \,^{\times }\left( \left. S\cap {\cal H}_{\pm }^2\right| _{%
{\Bbb R}^{+}}\right)  \label{1}
\end{equation}
are RHS.

Next, we assume that the free Hamiltonian, $K$, and the total Hamiltonian, $%
H=K+V$, have an absolutely continuous spectrum which is not degenerate and
coincides with the positive semiaxis. The asumption in the nondegeneracy of $%
K$ and $H$ is not strictly necassary and we insert it in order to simplify
the model and hence the notation. Under this circumstance, there is a
unitary operator between the Hilbert space of the free states of our system
and $\,L^2({\Bbb R}^{+})$, the space of the energy representation, that {\it %
diagonalizes} $K$, i.e., $UKU^{-1}$is the multiplication operator on $\,L^2(%
{\Bbb R}^{+})$. If ${\bf \Phi }_{\mp }=U^{-1}\,\left( \left. S\cap {\cal H}%
_{\pm }^2\right| _{{\Bbb R}^{+}}\right) $, we have two new RHS:

\begin{equation}
{\bf \Phi }_{\pm }\,\subset \,{\cal H\,\subset \,}^{\times }{\bf \Phi }_{\pm
}  \label{2}
\end{equation}
where ${\cal H}$ represents the total Hilbert space if $K$ has not
eigenvectors. If $K$ has eigenvectors, like in the Friedrichs model, we have
to consider the absolutely continuous part of ${\cal H}$ with respect to $K$%
, which is defined as: ${\cal H}_{\text{ac}}(K):=$ ${\cal H\ominus G}$. Here 
${\cal G}$ is the space of bound states for $K$.

Since the M\o ller operators, $\Omega _{\pm }$, exist and have the right
properties, we can define ${\bf \Phi }^{\pm }:=\Omega _{\pm }{\bf \Phi }%
_{\pm }$. We have new RHS:

\begin{equation}
{\bf \Phi }^{\pm }\,\subset \,{\cal H}_{\text{ac}}(H){\cal \,\subset \,}%
^{\times }{\bf \Phi }^{\pm }  \label{3}
\end{equation}
It is precisely in the space $^{\times }{\bf \Phi }^{\pm }$ where the Gamov
vectors live.

The purpose of this paper is to discuss some formulas and constructions that
sometimes appear in the literature \cite{12}, \cite{13}, \cite{14}, \cite{15}
and that find their own motivation precisely in the RHS formualtion of
resonance scattering. We shall present our arguments in the remaining part
of the present article.

\section{Presentation.}

As we have just mentioned in the Introduction, we wish to contribute to
clarify certain difficult aspects which remain obscure so far. We start with
a formula which is often showed and try to discuss its validity.

The point of departure is the RHS given in (\ref{3}) with plus sign. Let $%
\varphi ^{+}\in {\bf \Phi }^{+}$. The theorem of Gelfand and Maurin
establishes the existence of a complete set of generalized eigenvectors of
the total Hamiltonian $H$ such that, $\forall \,\varphi ^{+}\in {\bf \Phi }%
^{+}$

\begin{equation}
\varphi ^{+}=\int_0^\infty dE\,|E^{+}\rangle \,\langle ^{+}E|\varphi
^{+}\rangle  \label{4}
\end{equation}
and $H|E^{+}\rangle =E|E^{+}\rangle $. On the other hand, take $\psi ^{-}\in 
{\bf \Phi }^{-}$ and $\varphi ^{+}\in {\bf \Phi }^{+}$. We have the
following formula: \cite{1}, \cite{2}, \cite{5} 
\begin{equation}
(\psi ^{-},\varphi ^{+})=\int_0^\infty dE\,\langle \psi ^{-}|E^{-}\rangle
\,S(E+i0)\,\langle E^{+}|\varphi ^{+}\rangle  \label{5}
\end{equation}
where the functionals $|E^{-}\rangle \in \,^{\times }{\bf \Phi }^{-}$ are
also generalized eigenvectors of $H$. From (\ref{4}) and (\ref{5}), one
gets: 
\begin{equation}
\varphi ^{+}=\int_0^\infty dE\,|E^{-}\rangle \,S(E+i0)\,\langle
E^{+}|\varphi ^{+}\rangle  \label{6}
\end{equation}
Comparing (\ref{4}) with (\ref{6}), one is tempted to conclude that 
\begin{equation}
|E^{-}\rangle \,S(E+i0)=|E^{+}\rangle  \label{7}
\end{equation}
This is the first point we want to comment. First of all, we have to realize
that formulas (\ref{4}) and (\ref{6}) do not refer to the same kind of
object. On (\ref{4}), $\varphi ^{+}$ represents the unique element of $%
^{\times }{{\bf \Phi }^{+}}$ which is the image of $\varphi ^{+}\in {\bf %
\Phi }^{+}$ into $^{\times }{\bf \Phi }^{+}$ by the natural imbedding. On (%
\ref{6}), it is an element of $^{\times }{\bf \Phi }^{-}$ which does not
come from ${\bf \Phi }^{-}$ but from the Hilbert space (which contains ${%
{\bf \Phi }^{+}}$). Therefore, they are functionals on two different spaces.
In addition, the functionals $|E^{-}\rangle $ and $|E^{+}\rangle $ act on
two different spaces. Therefore, they cannot be proportional to each other.

Nevertheless, one may think that the functionals $|E^{-}\rangle $ and $%
|E^{+}\rangle $ could be extended into the space ${\bf \Phi }={\bf \Phi }%
^{+}+{\bf \Phi }^{-}$ and then finding a possible relation among them. The
space ${\bf \Phi }$ can be endowed with a topology using the topologies on $%
{\bf \Phi }^{+}$ and ${\bf \Phi }^{-}$.

In general, topologies on a locally convex space are given by a family of
seminorms. Seminorms have the same property of norms with the exception that
the seminorm of a nonzero vector may be zero. In particular, norms are
seminorms. If a vector space ${\bf \Phi }$ is the {\it direct} sum of two
locally convex spaces: ${\bf \Phi }={\bf \Phi }_1\bigoplus {\bf \Phi }_2$,
we can construct a locally convex topology on it defining seminorms. In fact
any $\varphi \in {\bf \Phi }$ can be uniquely written as $\varphi =\varphi
_1+\varphi _2$, where $\varphi _1\in {\bf \Phi }_1$ and $\varphi _2\in {\bf %
\Phi }_2$. Then, if $p_1$ is a seminorm on ${\bf \Phi }_1$ and $p_2$ is a
seminorm on ${\bf \Phi }_2$, $p(\varphi )=p_1(\varphi _1)+p_2(\varphi _2)$
defines a seminorm on ${\bf \Phi }$ and all seminorms on ${\bf \Phi }$ are
defined in this manner.

However, if ${\bf \Phi }_1\cap {\bf \Phi }_2$ is nontrivial, a vector of $%
{\bf \Phi }$ cannot be written in a {\bf unique} manner as the sum of a
vector of ${\bf \Phi }_1$ plus a vector of ${\bf \Phi }_2$, but there are
infinitely many choices in general. For this reason, the seminorms in the
sum space ${\bf \Phi }$ cannot be written as above. It is nevertheless true
that for a seminorm $p_1$ on ${\bf \Phi }_1$ and a seminorm $p_2$ on ${\bf %
\Phi }_2$ we have a unique seminorm $p$ on ${\bf \Phi }$, which is defined
as: 
\begin{equation}
p_{12}(\varphi )=\inf \{p_1(\varphi _1)+p_2(\varphi _2)\}  \label{8}
\end{equation}
where the infimum is taken over all possible forms of decomposing $\varphi
\in {\bf \Phi }$ as a sum of a vector $\varphi _1$ in ${\bf \Phi }_1$ and a
vector $\varphi _2$ in ${\bf \Phi }_2$.

In our case, this nonuniqueness is a bad property, and contributes to make
useless the space ${\bf \Phi }:={\bf \Phi }^{+}+{\bf \Phi }^{-}$. Let us
show why. One may think that the vectors $|E^{+}\rangle $ and $|E^{-}\rangle 
$ could be extended to this sum and then, they could be compared. This is
false for simple algebraic reasons. The problem arises because the
operations that carries ${\bf \Phi }^{+}$ and ${\bf \Phi }^{-}$ to spaces of
Hardy functions are different. This operation is a product of a M{\o }ller
operator times a unitary operator and we use a different M{\o }ller operator
in each case. Thus, if $\varphi =\varphi ^{+}+\varphi ^{-}$ and $\varphi
^{-}(E)$ is the Hardy function corresponding to $\varphi ^{+}$ and $\varphi
^{+}(E)$ the Hardy function corresponding to $\varphi ^{-}$, the function $%
\varphi (E)=\varphi ^{-}(E)+\varphi ^{+}(E)$ is well defined although it
depends on the way that we decompose $\varphi $ as a sum of an element of $%
{\bf \Phi }^{+}$ and an element of ${\bf \Phi }^{-}$. To show it, let us
take $\varphi \in {\bf \Phi }^{+}\cap {\bf \Phi }^{-}.$ As a vector in ${\bf %
\Phi }^{+},$ $\varphi $ is represented by the function $\varphi
_{-}(E)=U\Omega _{+}^{-1}\varphi .$ As a vector in ${\bf \Phi }^{-},$ $%
\varphi $ is represented by the function $\varphi _{+}(E)=U\Omega
_{-}^{-1}\varphi ,$ which are different in general. Thus, the mapping given
by 
\begin{equation}
\varphi \longmapsto \varphi (E)  \label{9}
\end{equation}
is not well defined. Unfortunately, this is the only serious candidate to
extend simulteneously $|E^{+}\rangle $ and $|E^{-}\rangle $.

However, we can give meaning to formula (\ref{7}) on a smaller space. Let us
take the intersection ${\bf \Phi }^{+}\cap {\bf \Phi }^{-}$ and assume that
it is nontrivial. Take $\varphi \in {\bf \Phi }^{+}\cap {\bf \Phi }^{-}.$
This $\varphi $ can be viewed either as an element of ${\bf \Phi }^{+}$, and
then we call it $\varphi ^{+}$, or as an element of ${\bf \Phi }^{-}$, and
then, we call it $\varphi ^{-}.$ Let us consider the following pair of
functions: 
\begin{equation}
\langle E^{+}|\varphi ^{+}\rangle =U\Omega _{+}^{-1}\varphi \;;\;\langle
E^{-}|\varphi ^{-}\rangle =U\Omega _{-}^{-1}\varphi   \label{(9)-1}
\end{equation}

From (\ref{(9)-1}), we can see that: 
\begin{eqnarray}
\langle E^{+}|\varphi ^{+}\rangle  &=&U\Omega _{+}^{-1}\Omega
_{-}U^{-1}\langle E^{-}|\varphi ^{-}\rangle   \nonumber \\
&=&US^{-1}U^{-1}\langle E^{-}|\varphi ^{-}\rangle   \nonumber \\
&=&S^{*}(E+i0)\langle E^{-}|\varphi ^{-}\rangle   \label{(9)-2}
\end{eqnarray}

This implies that: 
\begin{equation}
\langle E^{+}|=S^{*}(E+i0)\langle E^{-}|\Leftrightarrow |E^{+}\rangle
=S(E+i0)|E^{-}\rangle   \label{(9)-3}
\end{equation}

\noindent but this formula is valid only when both terms of the identity act
on vectors of ${\bf \Phi }^{+}\cap {\bf \Phi }^{-}.$

A similar situation happens with the Gamov vectors. One candidate of the
extension of $|f_0\rangle $ to ${\bf \Phi }$ is the following: 
\begin{equation}
\varphi =\varphi ^{+}+\varphi ^{-}\hskip0.5cm\text{then}\hskip0.5cm\langle
\varphi |f_0\rangle :=\langle \varphi ^{-}|f_0\rangle  \label{10}
\end{equation}
Since the decomposition is not unique, $|f_0\rangle $ is not well defined on 
$\varphi $.

Once we have raised the question whether there is a relation between $%
|E^{+}\rangle $ and $|E^{-}\rangle $ or not, it would be interesting to
provide an answer to this question. Take $E_0>0$. Consider the Schwartz
space $S$ and the distribution $\delta ^{*}(E-E_0)$ defined as 
\begin{equation}
\int_{-\infty }^\infty f(E)\,\delta ^{*}(E-E_0)\,dE=f^{*}(E_0)  \label{11}
\end{equation}
where $f^{*}(E_0)$ is the complex conjugate of the value of the Schwartz
function $f(E)$ at the point $E_0$. This distribution is an antilinear
continuous functional on $S$ and can be viewed as the complex conjugate of
the Dirac delta. Now, take the subspace given by the direct sum 
\begin{equation}
({\cal H}_{+}^2\cap S)\oplus ({\cal H}_{-}^2\cap S)  \label{12}
\end{equation}
This is a proper subspace of $S$ and therefore the functional $\delta
^{*}(E-E_0)$ can be restricted to it as it can be restricted to both ${\cal H%
}_{+}^2\cap S$ and ${\cal H}_{-}^2\cap S$. After the definitions of ${\bf %
\Phi }_{\pm }$ and ${\bf \Phi }^{\pm }$, there exists two unitary mappings $%
V_{\pm }$%
\begin{equation}
V_{\pm }=U\Omega _{\pm }^{-1}  \label{13}
\end{equation}
such that

\begin{equation}
V_{\pm }{\bf \Phi }^{\pm }=\left. {\cal H}_{\mp }^2\cap S\right| _{{\Bbb R}%
^{+}}=\theta _{\mp }\left( {\cal H}_{\mp }^2\cap S\right)  \label{14}
\end{equation}
where $\theta _{\mp }$ gives the relation between a function on ${\cal H}%
_{\mp }^2$ and its restriction to ${\Bbb R}^{+}$. This relation is one to
one, due to the van Winter theorem \cite{17}, \cite{5}.

By duality, we can extend all these operators to relations between the dual
spaces and the relations (\ref{13}) hold for the extended operators. In
particular:

\begin{equation}
V_{\pm }\,^{\times }{\bf \Phi }^{\pm }=U\Omega _{\pm }^{-1}\,^{\times }{\bf %
\Phi }^{\pm }=\,^{\times }\left( \left. {\cal H}_{\mp }^2\cap S\right| _{%
{\Bbb R}^{+}}\right) =\theta _{\mp }^{\times }\,\left[ ^{\times }\left( 
{\cal H}_{\mp }^2\cap S\right) \right]  \label{15}
\end{equation}

After these comments, we can see \cite{5} that for any fixed value $E_0\in 
{\Bbb R}$ and the definition of $|E_0^{\pm }\rangle $ ($\langle \varphi
^{\pm }|E_0^{\pm }\rangle =[\varphi ^{\mp }(E_0)]^{*}$ ; $\varphi ^{\mp
}(E)=\theta _{\mp }^{-1}V_{\pm }\varphi ^{\pm }$ ; $\forall \,\varphi ^{\pm
}\in {\bf \Phi }^{\pm }$) , that: 
\begin{equation}
|E_0^{+}\rangle =\Omega _{+}U^{-1}\theta _{-}^{\times }[\delta ^{*}(E-E_0)]
\label{16}
\end{equation}
\begin{equation}
|E_0^{-}\rangle =\Omega _{-}U^{-1}\theta _{+}^{\times }[\delta ^{*}(E-E_0)]
\label{17}
\end{equation}
This equations give: 
\begin{equation}
|E_0^{+}\rangle =\Omega _{+}U^{-1}\theta _{-}^{\times }(\theta _{+}^{\times
})^{-1}U\Omega _{-}^{-1}|E_0^{-}\rangle  \label{18}
\end{equation}

We observe that the operator that relates these two kets depends:

(i) On the {\it dynamics} through the M{\o }ller wave operators.

(ii.) On {\it causality} via the properties of Hardy functions through the $%
\theta _{\pm }$ operators. Nevertheless, its interesting to point out that
the relation between them is not of the type 
\[
|E^{+}\rangle =(\Omega _{-})^{-1}O\Omega _{+}|E^{-}\rangle 
\]
where $O$ is an operator, as suggested by (\ref{7}) (recall that $S=(\Omega
_{-})^{-1}\Omega _{+}$).

Summarizing, formula (\ref{7}) {\bf does not make sense }on the space ${\bf %
\Phi }={\bf \Phi }^{+}+{\bf \Phi }^{-}$

In any case, to make sense out of formula \ref{7}, we need to find a space
in which both $|E^{-}\rangle $ and $|E^{+}\rangle $ act. This will
necessarily imply a redefinition of either ${\bf \Phi }^{+}$ or ${\bf \Phi }%
^{-}$ or both. On the other hand, it seems natural that the space of
outgoing states be the result of the action of the $S$-operator on the space
of incoming states, so that:

\begin{equation}
S{\bf \Phi }_{+}={\bf \Phi }^{\text{out}}\Longleftrightarrow \Omega _{+}{\bf %
\Phi }_{+}=\Omega _{-}{\bf \Phi }^{\text{out}}  \label{19}
\end{equation}
where ${\bf \Phi }_{+}$ is defined as in \ref{2}. Here, $|E^{-}\rangle $ is
a functional on ${\bf \Phi }^{-}=\Omega _{-}{\bf \Phi }^{\text{out}}$ that
should be defined as in the standard case \cite{5}. Take $\varphi ^{-}\in
\Phi ^{-}$ and the function given by $\varphi _{+}(E)=U\Omega
_{-}^{-1}\varphi ^{-}$. This function is defined on ${\Bbb R}^{+}.$ Then, if 
$E_0>0,$ $\langle \varphi ^{-}|E_0^{-}\rangle $ is given by the value of the
function $[\varphi _{+}(E)]^{*}$ at the point $E_0.$ The meaning of $%
|E^{+}\rangle $ does not change as a functional on $\Omega _{+}{\bf \Phi }%
_{+}.$ Since these spaces are equal, $|E^{-}\rangle $ and $|E^{+}\rangle $
may be compared. To do it, let us choose $\varphi ^{\text{in}}\in {\bf \Phi }%
_{+}$ and $\psi ^{\text{out}}\in {\bf \Phi }^{\text{out}}.$ Then, $\psi
^{-}=\Omega _{-}\psi ^{\text{out}}$, $\varphi ^{+}=\Omega _{+}\varphi ^{%
\text{in}},$ we have:

\begin{eqnarray}
\langle \varphi ^{+}|E^{+}\rangle &=&\langle \Omega _{+}\varphi ^{\text{in}%
}|\Omega _{+}|E\rangle =\langle \varphi ^{\text{in}}|E\rangle =[\langle
E|\varphi ^{\text{in}}\rangle ]^{*}=  \nonumber \\
&=&[U\varphi ^{\text{in}}]^{*}(E)=[\varphi ^{-}(E)]^{*}\in \left. {\cal H}%
_{+}^2\cap S\right| _{{\Bbb R}^{+}}  \label{20}
\end{eqnarray}
where $K|E\rangle =|E\rangle $ for $E>0,$ (see discussion on the last part
of the paper on the Dirac kets $|E\rangle $ for the free Hamiltonian $K$.).
The action of $|E^{-}\rangle $ on an arbitrary $\varphi ^{+}\in {\bf \Phi }%
^{+}$ is defined as:

\begin{eqnarray}
\langle \varphi ^{+}|E^{-}\rangle &=&\langle \Omega _{+}\varphi ^{\text{in}%
}|\Omega _{-}|E\rangle =\langle \Omega _{-}^{\dagger }\Omega _{+}\varphi ^{%
\text{in}}|E\rangle =  \nonumber \\
&=&\langle S\varphi ^{\text{in}}|E\rangle =[S(E+i0)\varphi
^{-}(E)]^{*}=S^{*}(E+i0)\,\langle \varphi ^{+}|E^{+}\rangle  \label{21}
\end{eqnarray}

Thus,

\begin{equation}
|E^{-}\rangle =S^{*}(E+i0)\,|E^{+}\rangle \Longleftrightarrow
\,|E^{+}\rangle =S(E+i0)\,|E^{-}\rangle  \label{22}
\end{equation}
since $S(E+i0)$ for $E>0$ is a complex number with modulus equal to one.
This procedure has some inconvenients:

1.- The function:

\begin{equation}
\langle \psi ^{-}|E^{-}\rangle \,S(E+i0)\,\langle E^{+}|\varphi ^{+}\rangle
\label{23}
\end{equation}
is the function that appears under the integral sign in the explicit
expression for $(\psi ^{\text{out}},S\varphi ^{\text{in}}).$ This scalar
product is the point of departure of the construction of Gamov vectors ``a
la Bohm'' and the separation between the contribution of the Gamov vectors
and the background integral to the decaying process \cite{1}, \cite{5}. The
function (\ref{23}) is not meromorphic on the lower half plane, since $%
\langle \psi ^{-}|E^{-}\rangle \in \left. {\cal H}_{+}^2\cap S\right| _{%
{\Bbb R}^{*}}.$ To save this situation, we may choose $\psi ^{\text{out}}\in 
{\bf \Phi }_{+}\cap {\bf \Phi }_{-},$ so that $\langle \psi
^{-}|E^{-}\rangle \in \left. {\cal H}_{+}^2\cap S\right| _{{\Bbb R}^{+}}\cap
\left. {\cal H}_{-}^2\cap S\right| _{{\Bbb R}^{+}}.$ This intersection is
not trivial \cite{9}, but we do not know whether it is dense or not. Even if
this intersection were dense, this consideration will create problems to
define the time evolution for the Gamov vectors, as we shall see.

2.- Worst of all, the expression $(\psi ^{\text{out}},S\varphi ^{\text{in}})$
becomes totally useless for the theory of resonances. In fact, since $\psi ^{%
\text{out}}\in S{\bf \Phi }_{+},$ there exists $\psi ^{\text{in}}\in {\bf %
\Phi }_{+}$ such that $\psi ^{\text{out}}=S\psi ^{\text{in}}.$ Thus, $(\psi
^{\text{out}},S\varphi ^{\text{in}})=(S\psi ^{\text{in}},S\varphi ^{\text{in}%
})=(\psi ^{\text{in}},\varphi ^{\text{in}}),$ which is independent of the
scattering process and, therefore, does not show resonances.

We could pose the problem in other terms: Is it possible to write a formula
like

\begin{equation}
S\,|E^{-}\rangle =|E^{+}\rangle ?  \label{24}
\end{equation}

In order to do it, it seems clear that we should extend $S$ to $^{\times }%
{\bf \Phi }^{-}$ first. Let us analyze all possibilities. First of all, let
us take the vector space given by:

\begin{equation}
S\,{\bf \Phi }^{-}=\{\eta \in {\cal H}_{\text{ac}}(H)\;/\;\exists \,\varphi
^{-}\in {\bf \Phi }^{-}\;:\;\eta =S\varphi ^{-}\}  \label{25}
\end{equation}
where $S$ is the $S$-operator on ${\cal H}$. For simplicity, we shall assume
that $S{\cal H}_{\text{ac}}(H)={\cal H}_{\text{ac}}(H),$ which, under the
posed conditions, happens for instance if $H$ has no Hilbert space
eigenvalues.

Then, $S$ can be extended to the dual $^{\times }{\bf \Phi }^{-}$ by
duality, so that we have a new RHS:

\begin{equation}
S\,{\bf \Phi }^{-}\subset {\cal H}_{\text{ac}}(H)\subset \,^{\times }[S\,%
{\bf \Phi }^{-}]=S(^{\times }{\bf \Phi }^{-})  \label{26}
\end{equation}

In order to compare the functionals $S\,|E^{-}\rangle $ and $|E^{+}\rangle ,$
they should act on the same space. If $S\,|E^{-}\rangle $ acts on ${\bf \Phi 
}^{+},$ the bracket $\langle \varphi ^{+}|S|E^{-}\rangle $ must be well
defined for any $\varphi ^{+}\in {\bf \Phi }^{+}.$ This means that for all $%
\varphi ^{+}\in {\bf \Phi }^{+},$ $\exists \,\psi ^{-}\in {\bf \Phi }^{-}$
such that $S\psi ^{-}=\varphi ^{+},$ i.e., ${\bf \Phi }^{+}\subset S\,{\bf %
\Phi }^{-}.$ Also, if $|E^{+}\rangle $ acts on $S\,{\bf \Phi }^{-},$ for any 
$\psi ^{-}\in {\bf \Phi }^{-},$ then the bracket $\langle S\,\psi
^{-}|E^{+}\rangle $ makes sense. Thus, $S\psi ^{-}\in {\bf \Phi }%
^{+},\,\forall \,\psi ^{-}\in {\bf \Phi }^{-},$ which means that $S\,{\bf %
\Phi }^{-}\subset {\bf \Phi }^{+}.$ Consequently,

\begin{equation}
S\,{\bf \Phi }^{-}={\bf \Phi }^{+}  \label{27}
\end{equation}
This implies that for any $\varphi ^{+}\in {\bf \Phi }^{+}$ there exists $%
\psi ^{-}\in {\bf \Phi }^{-}$ such that $S\psi ^{-}=\varphi ^{+}$ and
viceversa. This identity has the following equivalent forms:

\begin{eqnarray}
\varphi ^{+} &=&S\,\psi ^{-}\Longleftrightarrow \Omega _{+}\varphi
_{+}=S\Omega _{-}\psi _{-}\Longleftrightarrow \Omega _{+}U^{-1}\varphi
_{-}(E)=  \nonumber \\
&=&S\Omega _{-}U^{-1}\psi _{+}(E)\Longleftrightarrow \varphi _{-}(E)=U\Omega
_{+}^{\dagger }S\Omega _{-}U^{-1}\psi _{+}(E)=  \nonumber \\
&=&U\Omega _{+}^{\dagger }U^{-1}USU^{-1}U\Omega _{-}U^{-1}\psi _{+}(E)
\label{28}
\end{eqnarray}
Since $USU^{-1}=S(E+i0),$ (\ref{28}) is equal to

\begin{equation}
\varphi _{-}(E)=S(E+i0)U\Omega _{+}^{\dagger }\Omega _{-}U^{-1}\psi
_{+}(E)=S(E+i0)US^{-1}U^{-1}\psi _{+}(E)  \label{29}
\end{equation}
Analogously, 
\[
US^{-1}U^{-1}\psi _{+}(E)=S^{*}(E+i0)\psi _{+}(E) 
\]
With the property that $S(E+i0),$ with $E>0,$ is a complex number with
modulus one, we finally have:

\begin{equation}
\langle E^{+}|\varphi ^{+}\rangle =\varphi _{-}(E)=\psi _{+}(E)=\langle
E^{-}|\psi ^{-}\rangle  \label{30}
\end{equation}

Formula $($\ref{30}) does not make sense if we keep for ${\bf \Phi }^{\pm }$
the definitions given in (\ref{13}). In fact, (\ref{30}) means that $V_{+}%
{\bf \Phi }^{+}=V_{-}{\bf \Phi }^{-},$ which is not true after (\ref{13}).
Therefore, in order to make formula (\ref{24}) mathematically meaningful, we
need to reconstruct ${\bf \Phi }^{\pm }$ with the mentioned property that $%
V_{+}{\bf \Phi }^{+}=V_{-}{\bf \Phi }^{-}.$ This is what we are going to do
in the next few paragraphs.

To begin with, let us consider now the space $S({\Bbb R}^{-})$ of all
Schwartz functions supported in the negative semiaxis ${\Bbb R}^{-}=(-\infty
,0].$ For any $f\in S({\Bbb R}^{-}),$ the function ${\frak H}f\pm if\in 
{\cal H}_{\pm }^2,$ where ${\frak H}$ is the Hilbert transform \cite{16}.

\medskip

{\bf Proposition}.- The Fourier transform ${\cal F\,(}{\frak H}f\pm if)$ is
the restriction to${\cal \ }{\Bbb R}^{\mp }$ of a Schwartz function.

\medskip

{\bf Proof}.- Being given a function $f$, its Hilbert transform ${\frak H}f$
represents the convolution of $f$ with the distribution ${\em PV}\left(
\frac 1x\right) $ (${\em PV}$ denotes Cauchy principal value). We know that:

\begin{equation}
2{\em PV}\left( \frac 1x\right) =\frac 1{x+i0}+\frac 1{x-i0}  \label{31}
\end{equation}
The Fourier transform of this distribution is given by

\begin{equation}
{\cal F}\left( {\em PV}\left( \frac 1x\right) \right) =i(H(-x)-H(x))=iG(x)
\label{32}
\end{equation}
where $H(x)$ is the Heaviside or step function. It is zero on the negative
semiaxis and one on the positive semiaxis. Since the Fourier transform of
the convolution of two functions is the product of the Fourier transforms of
both functions, we conclude that:

\begin{equation}
{\cal F\,(}{\frak H}f\pm if)=iG(x){\widehat{f}(x)}\pm i{\widehat{f}}(x)\hskip%
0.3cm\text{where\hskip0.3cm }{\widehat{f}}={\cal F}(f)  \label{33}
\end{equation}
This Fourier transform is explicitely given by given by

If $x>0$

\begin{equation}
{\cal F\,(}{\frak H}f+if)=0\hskip0.5cm;\hskip 0.5cm{\cal F\,(}{\frak H}%
f-if)=\left. -2i{\widehat{f}}(x)\right| _{{\Bbb R}^{+}}  \label{34}
\end{equation}

If $x<0$

\begin{equation}
{\cal F\,(}{\frak H}f+if)=\left. 2i{\widehat{f}}(x)\right| _{{\Bbb R}^{-}}%
\hskip0.5cm;\hskip 0.5cm{\cal F\,(}{\frak H}f-if)=0  \label{35}
\end{equation}
Equations $($\ref{34}) and (\ref{35}) give us the desired result.$%
\blacksquare $

\medskip

The functions which are restrictions of Schwartz functions to ${\Bbb R}^{\pm
}$ are dense in $L^2({\Bbb R}^{\pm }).$ After that, we conclude by the
Paley-Wienner theorem \cite{11} that 
\[
{\cal (}{\frak H}f\pm if)\,S({\Bbb R}^{-}) 
\]
is dense in ${\cal H}_{\pm }^2.$

As a consequence of the van Winter theorem \cite{17}, \cite{5}, the
restrictions of a function in ${\cal H}_{\pm }^2$ to either ${\Bbb R}^{+}$
or ${\Bbb R}^{-}$ determine all the values of the function. Furthermore,
these restrictions are dense in $L^2({\Bbb R}^{+})$ and $L^2({\Bbb R}^{-}).$
Since the functions in 
\[
{\cal (}{\frak H}f\pm if)\,S({\Bbb R}^{-}) 
\]
are dense in ${\cal H}_{\pm }^2,$ we immediately follow that their
restrictions to $L^2({\Bbb R}^{+})$ are dense in this space.

Furthermore, the restrictions of the functions in ${\cal (}{\frak H}f\pm
if)\,S({\Bbb R}^{-})$ to ${\Bbb R}^{+}$ satisfy an important property: they
admit a unique extension to ${\Bbb R}^{-}$ to be in ${\cal H}_{+}^2$ and
another unique extension to ${\Bbb R}^{-}$ to be in ${\cal H}_{-}^2$ and
these extensions are different, since the intersection ${\cal H}_{+}^2\cap 
{\cal H}_{-}^2=\{{\bf 0\}.}$ To see it, recall that any $f\in S({\Bbb R}%
^{-}) $ is zero on the positive semiaxis, so that $\left. {\frak H}f\pm
if\right| _{{\Bbb R}^{+}}=\left. {\frak H}f\right| _{{\Bbb R}^{+}}.$ Thus, $%
{\frak H}f\pm if\in {\cal H}_{\pm }^2$ but are indistinguisable on the
positive semiaxis.

Now, let us define the following space:

\begin{equation}
\Delta =\{g\in L^2({\Bbb R}^{+}),\;\text{such that }\exists \,f\in S({\Bbb R}%
^{-})\text{ with }g=\left. {\frak H}f\pm if\right| _{{\Bbb R}^{+}}\}
\label{36}
\end{equation}

As we have seen, $\Delta $ is dense in $L^2({\Bbb R}^{+}).$ The space of the
restrictions to ${\Bbb R}^{-}$ of Schwartz functions can be endowed with a
nuclear locally convex topology exactly as we do on the Schwartz space.
Then, it is immediate to show that:

\begin{equation}
\Delta \subset \,L^2({\Bbb R}^{+})\subset \,^{\times }\Delta  \label{37}
\end{equation}
is a RHS. If we construct:

\begin{equation}
{\bf \Phi }^{\pm }=V_{\pm }^{-1}\Delta  \label{38}
\end{equation}
we have triplets with the required conditions so that the formula $%
S\,|E^{-}\rangle =|E^{+}\rangle $ is correctly defined. Note that in this
case, we have for the spaces defined right before (\ref{2}): ${\bf \Phi }%
_{+}={\bf \Phi }_{-}=U^{-1}\Delta .$

It seems that we have found triplets that satisfy all possible good
properties. This is however false! Certainly, we can define the Gamow
vectors using the triplet (\ref{3}) with ${\bf \Phi }^{\pm }$ defined as in (%
\ref{38}), but this causes a severe dificulty: if $\varphi (E)\in \Delta ,$
then, $e^{itE}\varphi (E)\notin \Delta $ for any $t\neq 0.$ The reason is
very simple: assume, for instance, that $t>0.$ For $f(E)\in \Delta ,$ the
integrals:

\begin{equation}
\int_{-\infty }^\infty |e^{it(E+i\alpha )}\,f(E+i\alpha )|^2\,dE=e^{-t\alpha
}\,\int_{-\infty }^\infty |f(E+i\alpha )|^2\,dE  \label{39}
\end{equation}
are uniformly bounded for $\alpha >0,$ but they {\bf are not} for $\alpha
<0. $ This means that there is {\bf no time evolution} for ${\bf \Phi }^{\pm
}$ and its duals, when they are defined as in (\ref{38}), because then, for
any $t\neq 0$ and any $\varphi ^{\pm }\in {\bf \Phi }^{\pm },$ $%
e^{itH}\varphi ^{\pm }\notin {\bf \Phi }^{\pm }.$

{\bf Summarizing}: If we want to give meaning to formula (\ref{7}), we loose
information about $S$-matrix poles. If we want to give meaning to formula (%
\ref{24}), we loose the time behaviour for Gamov vectors. $\blacksquare $

We want to end this paper with another comment. Let us consider again the
spaces : ${\bf \Phi }_{\pm }$ and the RHS given in (\ref{2}). We know that
the free Hamiltonian $K$ reduces these spaces ($K{\bf \Phi }_{\pm }\subset 
{\bf \Phi }_{\pm }$) and is continuous on them. Therefore, it can be
continuously extended to ${\bf \Phi }_{\pm }^{\times }$ by duality. Since
the spectrum of $K$ coincides with the positive semiaxis ${\Bbb R^{+}}$, for
any $E\in {\Bbb R^{+}}$ there is a generalized eigenvector $|E_{\pm }\rangle 
$ of $K$ in ${\bf \Phi }_{\pm }^{\times }$: $K|E_{\pm }\rangle =E|E_{\pm
}\rangle $. These functionals are defined as follows: Let $\varphi _{\pm
}\in {\bf \Phi }_{\pm }$ and $U\varphi _{\pm }=\varphi _{\mp }(E)$. Take $%
E^{\prime }\in {\Bbb R^{+}}$, the functional $|E_{\pm }^{\prime }\rangle $
is defined by the mapping: 
\begin{equation}
\varphi _{\pm }\longmapsto [\varphi _{\mp }(E^{\prime })]^{*}=\langle
\varphi _{\pm }|E_{\pm }^{\prime }\rangle  \label{40}
\end{equation}

We want to show that $|E_{\pm }\rangle $ are indeed the restrictions to $%
{\bf \Phi }_{\pm }$ of a functional on a bigger space. This bigger space is: 
\begin{equation}
{\bf \Psi }={\bf \Phi }_{+}+{\bf \Phi }_{-}  \label{41}
\end{equation}

Let us take $\varphi \in {\bf \Psi }$. Then, $\varphi =\varphi _{+}+\varphi
_{-}$ and this decomposition is not unique. Let us define the action of $U$
on $\varphi \in {\bf \Psi }$ as $\varphi (E):=U\varphi =U\varphi
_{+}+U\varphi _{-}=\varphi _{-}(E)+\varphi _{+}(E)$. Obviously, this mapping
is well defined and does not depend on the decomposition of $\varphi $. For
any $E\in {\Bbb R^{+}}$, the mapping on ${\bf \Psi }$ given by $\varphi
\longmapsto [\varphi (E)]^{*}=\langle \varphi |E\rangle $ is well defined.
To prove its continuity, we write: 
\[
|\varphi (E)|\le |\varphi _{-}(E)|+|\varphi _{+}(E)|\le \sup \{|\varphi
_{-}(E)|\}+\sup \{|\varphi _{+}(E)|\}= 
\]
\begin{equation}
=p_0(\varphi _{+})+p_0(\varphi _{-})  \label{42}
\end{equation}
Here, $p_0$ represents one of the seminorms (here are indeed norms) that
provide the topology on $\Phi _{\pm }$ \cite{18}. Since the above inequality
hold for any decomposition of $\varphi $, we finally have: 
\begin{equation}
|\varphi (E)|\le \inf \{p_0(\varphi _{+})+p_0(\varphi _{-})\}  \label{43}
\end{equation}
proven the desired continuity. The restriction of the functional $|E\rangle $
to ${\bf \Phi }_{\pm }$ is $|E_{\pm }\rangle $. This justifies the unique
notation as $|E\rangle $ for these two functionals that appear in some
publications \cite{1}.

\bigskip

{\bf Acknowledgements}.- We are grateful to Profs A. Bohm, M. Castagnino, R.
Laura and Mr. D. Arb\'{o} for enlightening discussions. Partial finantial
support is acknowledged to the UE contract number CI1*-CT94-0004, to the CO
1/196 of the ''Junta de Castilla y Le\'{o}n'' Proyect and the Intercampus
Programme, and also to a Grant from the ''Foundation pour la Recherche
Foundamentale OLAM''.

\end{document}